\documentclass[epj]{svjour}

\def\be{\begin{equation}}
\def\ee{\end{equation}}
\def\bea{\begin{eqnarray}}
\def\eea{\end{eqnarray}}
\def\nn{\nonumber}                        
\def\ket#1{\hbox{$\vert #1\rangle$}}   
\def\bra#1{\hbox{$\langle #1\vert$}}   
\def\bold#1{\setbox0=\hbox{$#1$}%
      \kern-.02em\copy0\kern-\wd0
      \kern.04em\copy0\kern-\wd0
      \kern-.02em\raise.0433em\box0 }
\def\smbold#1{\setbox0=\hbox{$\scriptstyle#1$}%
      \kern-.02em\copy0\kern-\wd0
      \kern.04em\copy0\kern-\wd0
      \kern-.02em\raise.0433em\box0 }

\usepackage{graphics}
\usepackage{graphicx}
\usepackage{epsfig}
\begin{document}
\title{Nucleon electroweak form factors in a meson-cloud model}
\author{B. Pasquini\inst{} \and S. Boffi\inst{}}
\institute{Dipartimento di Fisica Nucleare e Teorica, Universit\`a degli
Studi di Pavia and INFN, Sezione di Pavia, Pavia, Italy}
\date{Received: date / Revised version: date}
%
\abstract{
The meson-cloud model of the nucleon consisting of a system of three valence 
quarks surrounded by a meson cloud is applied to study the electroweak 
structure of the proton and neutron. 
The electroweak nucleon form factors are calculated within a light-front
approach, 
by obtaining an overall good description of the experimental data.
Charge densities as a function of the transverse distance with 
respect to the direction of the three-momentum transfer are also discussed.
} 
\maketitle
%


\section{Introduction}
\label{sect:intro}

The relevant role played by the meson cloud of the nucleon is clearly
manifest in the study of the low-energy electromagnetic structure of the 
nucleon. 
Here we want to consider the role of the meson-cloud in the calculation
of the electroweak form factors of the nucleon within
a relativistic light-front approach
 where the  nucleon light-cone wave function (LCWF)  is derived by performing 
a baryon-meson Fock-state expansion. In the one-meson approximation
 the nucleon state 
$\ket{\tilde N}$ is pictured as being part of the time a bare 
nucleon, $\ket{N}$, and part of the time a baryon-meson system, $\ket{BM}$. 
The bare nucleon is formed by three valence quarks identified as constituent
 quarks, while the baryon-meson system is 
assumed to include configurations with the baryon being a nucleon or a Delta
 and the accompanying meson being a pion as well as a vector meson such as the
 $\rho$ or
 the $\omega$.
The calculation of the 
electroweak form factors of the proton and neutron is illustrated in 
Sect.~\ref{sect:ff}, and the results are presented and discussed in 
Sect.~\ref{sect:results}. 


\section{Electroweak form factors of the nucleon in the meson-cloud model}
\label{sect:ff}

The problem of considering the meson cloud surrounding a system of three 
valence quarks has been addressed already in the past in a variety of 
papers~\cite{Speth98,DHSS97,FGLP06,Julia06a,Giannini07}, 
and it has recently been revisited to study the generalized parton 
distributions~\cite{PB06} and the electroweak form factors of the 
nucleon~\cite{PB07}.
We refer to these last two works for a detailed explanation of the derivation
of the LCWF of the nucleon in the meson-cloud model 
in terms of a bare (three quark)   and
a baryon-meson (five quark) component, while in the following we 
review the main steps of the convolution formalism for the calculation
 of the electromagnetic and axial form factors of the nucleon.

The Dirac ($F_1$) and Pauli ($F_2$) form factors of the nucleon 
are given by the spin conserving and the spin-flip matrix elements of the 
vector current $J_V^+=J_V^0+J_V^3$ 
\bea
F_1(Q^2)&=&\langle \tilde p+\tilde q, \frac{1}{2}|J^+_V|\tilde p,\frac{1}{2}
\rangle,\\
(q_x+iq_y)F_2(Q^2)&=&2 M_N 
\langle \tilde p+\tilde q, -\frac{1}{2}|J^+_V|\tilde p,\frac{1}{2}\rangle,
\eea
where $q$ is the momentum transferred, and $Q^2=-q^2$.
The corresponding relation between the axial current $J_A$ and the axial form factor $G_A(Q^2)$ is
\bea
G_A(Q^2)&=&
\langle \tilde p+\tilde q, \frac{1}{2}|J^+_A|\tilde p,\frac{1}{2}
\rangle.
\eea
The convolution formalism derived in the following formulas
 apply for the electomagnetic as well as for the axial current, 
and we will use $J^+$ for either one of the two currents.
The calculation of the 
form factors
is conveniently done in a coordinate frame with $q^+=0$, where 
the current  matrix elements can be computed as a simple overlap of
 Fock-space wave functions with the same number of partons, 
with all off-diagonal terms involving pair
 production or annihilation by the current or vacuum vanishing. In the 
present meson-cloud model, we need to consider the contributions from the 
diagonal overlap between the bare-nucleon state, on one side, and the $BM$ 
components, on the other side. Furthermore, the  current is a 
sum of one-body currents, $J^+=\sum_{B,M} J^+_B+J^+_M$,
 which involves individual hadrons one at a time. 
This corresponds  to assuming that there are no interactions among the 
particles 
in a multiparticle Fock state during the interaction with the photon. 
Therefore the external probe can scatter either on the bare nucleon,
$|N\rangle$, or one of the constituents  of the higher Fock states, 
$|BM\rangle$.
As a result, the matrix elements of the current can be 
written as the sum of the following two contributions
\begin{eqnarray}
& &\bra{\tilde p'_N,\lambda'_N,\tilde N} 
J^+_{V}\ket{\tilde p_N,\lambda_N,\tilde N}=
Z\,I^N_{\lambda'_N,\lambda_N}+
\delta I_{\lambda'_N,\lambda_N}.
\label{eq:current}
\end{eqnarray}
In Eq.~(\ref{eq:current}) $I^N$ is the contribution from the bare nucleon,
$Z$ is the probability of finding a bare nucleon in the physical nucleon,
and $\delta I$ 
is the contribution from the $BM$ Fock components of the physical nucleon.
This last term can further be split into two contributions, with the active
particle being the baryon ($\delta I^{(B'B)M}$) or the meson  
($\delta I^{(M'M)B}$), i.e.
\begin{eqnarray}
\delta I_{\lambda'_N,\lambda_N}=
\sum_{B,B',M} \,\delta I^{(B'B)M}_{\lambda'_N,\lambda_N}
+\sum_{M',M,B} \, \delta I^{(M'M)B}_{\lambda'_N,\lambda_N}.
\label{eq:ff_bm}
\end{eqnarray}
The two terms in Eq.~(\ref{eq:ff_bm}) can explicitly be obtained 
by folding the current matrix elements of the baryon and meson
 constituents with the probability amplitudes describing the distributions of 
these constituents in the dressed initial and final nucleon. 
Furthermore, by using the kinematical nature of the light-front boost,
the baryon and meson current matrix elements can be factor out of the internal momentum integration, and the final results read
\bea
& &
\delta I^{(B'B)M}_{\lambda'_N,\lambda_N}(Q^2)
= 
\sum_{\lambda,\lambda',\lambda''}
I^{B'B}_{\lambda'\lambda}(Q^2)\int
{\rm d}y_B
\int
\frac{{\rm d}^2{\mathbf p}_{B\perp}}{2(2\pi)^3}\nn\\
& &\times
\phi^{\lambda_N\,(N,BM)}_{\lambda''\lambda}
(y_B,{\mathbf k}_{B\perp})
[\phi^{\lambda'_N\,(N,B'M)}_{\lambda''\lambda'}
(y_B,{\mathbf k}'_{B'\,\perp})]^*,
\nonumber\\
& &
\label{eq:I_B2}
\\
& &
\delta I^{(M'M)B}_{\lambda'_N,\lambda_N}(Q^2)
=
\sum_{\lambda,\lambda',\lambda''}
I^{M'M}_{\lambda'\lambda}(Q^2)\nn\\
& &\times\int
{\rm d}y_M
\int
\frac{{\rm d}^2
{\mathbf p}_{M\perp}}{2(2\pi)^3}
\phi^{\lambda_N\,(N,BM)}_{\lambda''\lambda}(1-y_M,-{\mathbf k}_{M\perp})\nn\\
& &\times
[
\phi^{\lambda'_N\,(N,BM')}_{\lambda''\lambda'}(1-y_M,-{\mathbf k}'_{M'\perp})
]^*,
\label{eq:I_M2}
\end{eqnarray}
where ${\mathbf k}_{B(M)\perp}={\mathbf p}_{B(M)\perp}-(1-y_{B(M)})
{\mathbf q}_{\perp}/2$ and ${\mathbf k}'_{B(M)\perp}={\mathbf p}_{B(M)\perp}+(1-y_{B(M)})
{\mathbf q}_{\perp}/2.$
In Eqs.~(\ref{eq:I_B2}) and (\ref{eq:I_M2}), the function
$\phi^{\lambda_N\,(NB,M)}_{\lambda''\lambda'}(y,{\mathbf k}_{\perp})$
is the probability amplitude to find a physical nucleon with helicity 
$\lambda_N$ in a state consisting of a virtual baryon $B$ and a virtual 
meson $M$, with the baryon having helicity $\lambda''$, longitudinal momentum 
fraction $y$ and transverse momentum ${\mathbf k}_{\perp}$, and the meson  
having helicity $\lambda'$, longitudinal momentum fraction $1-y$ and transverse
 momentum $-{\mathbf k}_{\perp}$.
In Eq.~(\ref{eq:ff_bm}) the sum over all the possible $BM$ 
configurations leads to contributions from both the diagonal
 current matrix elements with the same hadrons in the initial and final state
($B'=B$  and $M'=M$ in Eq.~(\ref{eq:I_B2}) and (\ref{eq:I_M2}), respectively),
and the current matrix elements involving the  transition
between different hadron states 
(i.e. the terms with $B'\neq B$ and $M'\neq M$ 
in Eq.~(\ref{eq:I_B2}) and (\ref{eq:I_M2}), respectively).
Finally, the current matrix elements between the bare-hadron states are
 calculated as overlap integrals of the hadron LCWFs, corresponding to 
the minimal Fock-state composition of valence quarks.  

\section{Results and discussion}
\label{sect:results}

The instant-form wave function of the bare-hadron states is constructed as a 
product of a momentum wave function, which is spherically symmetric and 
invariant under permutations, and a spin-isospin wave function, which is 
uniquely determined by SU(6) symmetry requirements. 
The transformation to the light-cone form is performed via Melosh rotations 
which break the SU(6) symmetry and introduce
components with orbital angular momentum different from zero, 
as explained in Ref.~\cite{PB07,BPT03}.
The functional form of the momentum wave functions is given in Ref.~\cite{PB07}, 
\begin{figure}[b]
\includegraphics[width=9.6 cm]{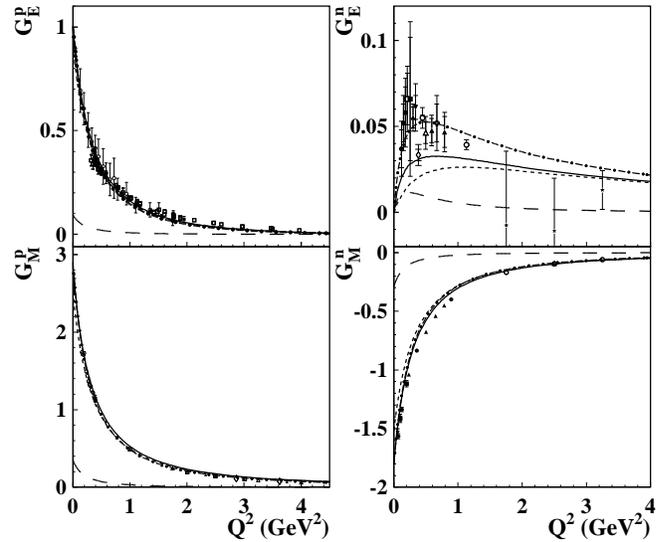}
\caption{ The four nucleon electromagnetic form factors compared with 
the world data.
considered in the analysis of Ref.~\cite{FW03} and the recent
 JLab data~\cite{Punjabi05} using 
$G_E^p=(\mu^p G_E^p/G_M^p)/(1+Q^2/0.71{ \rm GeV}^2)^2$ (open squares).
 Dotted curve for the contribution of the meson cloud; dashed curve
for the valence-quark contribution 
with SU(6) instant-form nucleon wave function; solid curve for the
sum of the two contributions;
 dashed-dotted curve for the total result 
with 1\% mixed-symmetry  $S'$-state in the bare nucleon wave function.
}
\label{fig:fig1}
\end{figure}
and it depends on three free parameters fitted to the proton ($\mu^p$) and 
neutron ($\mu^n$) anomalous magnetic moments, the axial coupling constant 
of the proton $g_A^p=G_A^p(0)$,  and the nucleon 
electromagnetic form factors at $Q^2=0.15$ and $0.45$ GeV$^2$.
The fitting procedure is performed by allowing  
a $5\%$ uncertainty.
For the anomalous magnetic moments we find $\mu^p=2.87$ and $\mu^n=-1.80,$ 
with a contribution from the meson cloud equal to $12\%$ and $ 15\%$, 
respectively,
while the result for the axial coupling constant is $g_A^p=1.20$ with a smaller 
contribution from the meson cloud of about $6\%$. 
These results are pretty close to the experimental values $\mu^p=2.79$, $\mu^n=-1.913$ and $g_A^p=1.267$, and especially for $\mu^n$ represent a significant 
improvement
with respect to constituent quark model calculations with only the 
contribution from the three valence quarks~\cite{Julia04,Graz01,cardarelli}.

In Fig.~\ref{fig:fig1} the results for the magnetic 
$G_M=F_1 - Q^2/(4M_N^2) F_2$ and electric $G_E= F_1+F_2$ form factors are shown
in comparison with  the experimental data.
 A rather good fit is obtained in the proton case in the whole range of
 available data, while in the neutron case the fit is less satisfactory. 
In
 any case, the contribution from the meson cloud is smooth and only 
significant for $Q^2< 0.5$ GeV$^2$ with a maximum at $Q^2=0$. 
In order to improve the result for the neutron electric form factor we also 
included a $1\%$ admixture  of mixed-symmetry $S'$-wave components, along the 
lines of Ref.~\cite{Julia04}. 
The contribution from such mixed-symmetric 
component is hardly visible in the case of the proton form factors and the 
neutron magnetic form factor, while it appreciably improves the agreement with 
the experimental data in the case of $G_E^n$.
\noindent
\begin{figure}[t]
\includegraphics[width=9.6 cm]{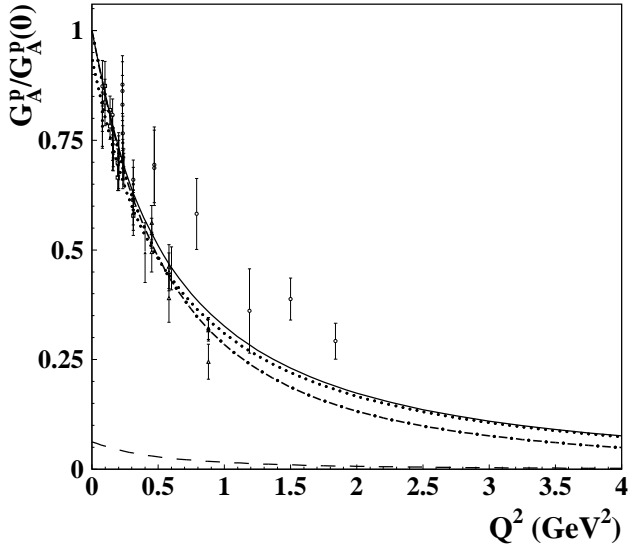}
\caption{The axial  form factor of the proton. Dotted (dashed) curve for the contribution of the meson cloud (valence quarks). Solid curve for the total result. Dot-dashed curve for the phenomenological dipole form. 
Data points are the world data considered in Ref.~\cite{BEM02}.}
\label{fig:fig5}
\end{figure}

The predicted axial form factor of the proton is shown in Fig.~\ref{fig:fig5}. 
The meson-cloud contribution is only significant at low 
values of $Q^2$, and the observed dipole form of the axial form factor,
 i.e. $G_A^p(Q^2)/G_A^p(0)=1/(1+Q^2/M_A^2)^2$ with $M_A =1.069$ GeV, 
is well reproduced. 
\begin{figure}[t]
\includegraphics[width=9.6 cm]{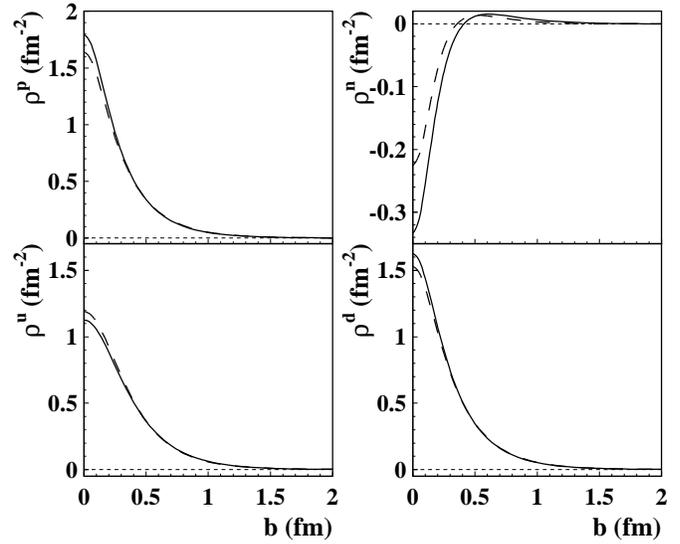}
\caption{
Upper panels: The proton (left) and neutron (right) charge density 
as a function of the impact parameter $b$. 
Lower panels: The up (left) and down (right) transverse charge densities in the neutron.
Solid curves for a SU(6)-symmetric instant-form wave function, 
dashed curves with mixed-symmetry components.}
\label{fig:fig8}
\end{figure}
\newline
The two-dimensional Fourier transform of $F_1$ in the transverse plane
perpendicular to the direction of motion of the nucleon allows us to map
the charge density $\rho(b)$ of the nucleon as function
of the transverse distance $b$
from the nucleon center.
As it was first pointed out in Ref.~\cite{Miller07}, 
the main advantage of the two-dimensional Fourier transform is that we 
can have a probabilistic interpretation for $\rho(b)$, 
at variance with the three dimensional Fourier transform of Sachs 
form factors.
The corresponding nucleon densities
 are plotted in the upper panels of Fig.~\ref{fig:fig8}.
While the negative tail of the neutron distribution has the well known interpretation in terms of the pion cloud, the negative charge density near the origin
appears to be mysterious. An intuitive  understanding of this result has been 
recently given in Ref.~\cite{Burkardt}, suggesting that u quark in the neutron have 
a  larger p-wave component than d quarks, being therefore suppressed at the
origin as shown in the lower panels of Fig.~\ref{fig:fig8}.
When considering also the Fourier transform of the axial form factor, one can define densities for longitudinally polarized quark in a polarized
nucleon~\cite{PB07,PB07b}, finding that the positive helicity up 
quarks in the proton are preferentially aligned with the proton helicity, 
while the opposite occurs for down quarks. 



\end{document}